\documentclass{PoS}
\usepackage[authoryear,square]{natbib}
\bibpunct{(}{)}{;}{a}{}{,}

\title{The quest for extragalactic magnetic fields}
\ShortTitle{Extragalactic magnetic fields}
\author{\speaker{Franco Vazza}$^{1,2}$
\\
	$^1$Hamburg University (Hamburg Observatory), Gojenbergsweg 112, 21029,
Germany;
    E-mail:\email{franco.vazza@hs.uni-hamburg.de}
 }

\newcommand{\enzo}{\it {\small ENZO}}

\abstract{{\bf Abstract} We review the observational and theoretical constraints on extragalactic magnetic fields across cosmic environment. In the next decade, the combination of sophisticated numerical simulations and various observational probes might succeed in constraining the still elusive origin of magnetic fields on the largest scales in the Universe.}

\FullConference{Neutrino Oscillation Workshop,\\
                4 - 11 September, 2016,\\ 
                Otranto (Lecce, Italy)}

\begin{document}

\section{Introduction - The mystery of magnetogenesis}
Understanding the origin of observed magnetic fields on $\sim \rm Mpc$ scales is still a challenge. While we roughly understand how primordial energy fluctuations of the cosmic microwave background (CMB) originated cosmic structures, we have yet no clear view of the origin of extragalactic magnetic fields \citep[][]{sub16}. 
Were the needed weak (and yet undetected) seed magnetic fields already present at the CMB, or were they only later released by forming galaxies? 
The observed properties of magnetic fields in galaxy clusters can roughly be reproduced by simulations, starting eiter from primordial fields or through the magnetisation from galactic winds and jets \citep[e.g.][and references therein]{donn09}.
On the other hand, the distribution of magnetic fields in cluster outskirts and in filaments is yet poorly constrained from observations and simulations, and there we expect
the different scenarios of magnetogenesis to diverge, owing to the weaker (or absent) level of dynamo amplification \citep[e.g.][]{va14mhd}.  Future observations should be able to kill alternative scenarios, by constraining the real distribution of magnetic fields across scales. In this contribution, we give a short overview of possible future ways to tackle this challenge, based on preliminary results of our ongoing campaign of large magneto-hydrodynamical (MHD) simulations of cosmic magnetism with {\enzo} \citep[][]{enzo13}.

\section{Cosmological MHD simulations}
\begin{figure}
 \includegraphics[width=0.99\textwidth,height=0.25\textwidth]{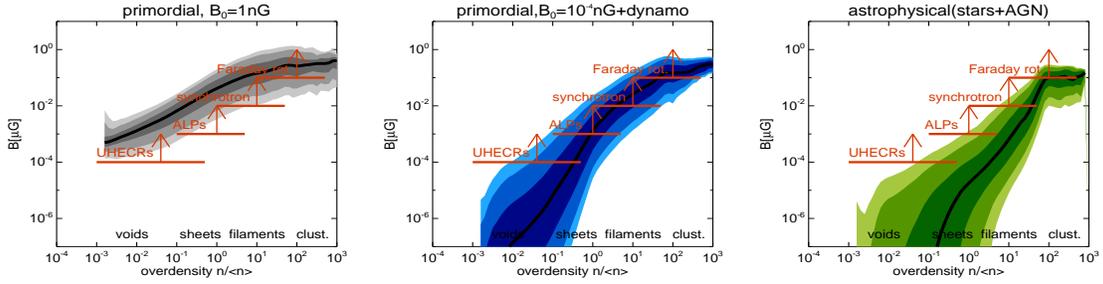}
\caption{Magnetic field strength as a function of gas overdensity in different seeding models. The colors mark the 5th,10th,25th, 50th, 75th, 90th and 95th percentile of the distributions. The additional red arrows mark the observational constraints discussed in the main text.}
\label{fig:models}
\end{figure}

Figure \ref{fig:models} shows the predicted distribution of simulated magnetic field strength across different bins in gas overdensity, for 3 of our {\enzo} resimulations of a $85^3 \rm Mpc^3$ volume using $1024^3$ cells/dark matter particles, metal-dependent gas cooling, feedback from star formation and active galactic nuclei (simulated using black-hole sink particles), and alternatively including: a) a primoridal uniform field of $B_0=1 ~\rm nG$ (comoving); b) a primordial uniform field of $B_0=10^{-4} ~\rm nG$ (comoving) with a run-time sub-grid modelling of dynamo amplification from solenoidal motions; c) magnetic seeding exclusively from star and AGN feedback, assuming a $\sim 10\%$ conversion of feedback energy into magnetic energy. Magnetic fields in high density regions are very similar in all models, yet the trends diverge in lower density regions. The same figure also shows where different observations should be sensitive to extragalactic magnetic fields (see below).

\section{Observational techniques and constaints}

Relativistic electrons accelerated by strong shocks in the cosmic web should emit continuum and polarised radio emission  \citep[e.g.][]{2011JApA...32..577B}. SKA-LOW
and its pathfinders and precurs may detect the tip of the iceberg of such emission, provided that the magnetic fields are $\geq 10 ~\rm nG $  \citep[][]{va15ska,va15survey}. Detections (or even robust upper limits) will constrain: a) the combination of magnetic fields  and electron acceleration efficiency at shocks, $\propto B^2 \times \xi_e$; b) the distribution of magnetic fields on $\geq 10~\rm Mpc$, yielding fundamental clues on their origin \citep[][]{va16radio}; c) the magnetic field obliquity at shocks accelerating radio emitting electrons \citep[][]{wi16}.

Additionally, future large surveys in polarisation (e.g. with ASKAP, MEERKAT and SKA-MID) will probe the topology of extragalactic fields through  Faraday Rotation ($\propto B_{\rm par} n_e$, where $B_{\rm par}$ is the magnetic field component along the line of sight and $n_e$ is the electron density), either with long exposures \citep[][]{2015arXiv150100321B}, or statistical techniques \citep[][]{2014ApJ...790..123A}. Our simulations suggest that detections should be feasible for $\geq 0.1 ~\rm \mu G$ fields in cluster outskirts.

The arrival direction at Earth of ultra-high energy cosmic rays (UHECRs) carries information in the large-scale distribution of magnetic fields \citep[e.g.][]{2005JCAP...01..009D}. 
With recent simulations, we showed that the distribution of UHECRs is sensitive to the magnetisation of voids. The simulated angular distribution of UHECRs gets 
too anistropic compared to available observational constraints, if the magnetisation of voids exceeds $\sim 0.1 ~\rm nG$ \citep[][]{hack16}.

Finally, extragalactic magnetic fields may also cause the oscillation of high-energy photons into axion-like particles (ALPs), a possibility that has been suggested to explain the lack of absorption in the spectra of distant blazars \citep{2012PhRvD..86g5024H}. In recent work \citep[][]{mo16} we simulated the propagation of ALPs in our MHD runs, finiding that significant photon-ALPs conversions are produced in lines of sight crossing structures with $\sim 1-10 ~\rm nG$ on scales of a few $\sim \rm Mpc$, i.e. in filaments and cosmic sheets along the line of sight of high-z sources.

\section{Conclusion}
In the near future the puzzle of magnetogenesis may be solved by combining complex simulations and observations of cosmic magnetic fields on different scales.

\section*{Acknowledgments}
These computations were produced on Piz Daint (ETHZ-CSCS, Lugano) under allocation s585. We acknowledges financial support from the grant VA 876-3/1 by DFG.

\bibliographystyle{apj}
\bibliography{franco}

\end{document}